\documentclass[journal,10pt,onecolumn]{IEEEtran}
\usepackage[table]{xcolor}
\usepackage{times}
\usepackage{epsfig}
\usepackage{amsmath}
\usepackage{amsthm}
\usepackage{amsfonts}
\usepackage{graphicx}
\usepackage{amssymb}
\usepackage{amstext}
\usepackage{latexsym}
\usepackage{color,colortbl}
\usepackage{ifthen}
\usepackage{multirow}
\usepackage{verbatim}
\usepackage{array,tabularx}
\usepackage{arydshln}
\usepackage[mathscr]{euscript}
\usepackage{accents}
 \usepackage{cite}
\usepackage{hhline}
\usepackage{caption}
\usepackage{subcaption}
\usepackage{enumerate}
\usepackage{xcolor}
\usepackage{mathtools}
\usepackage{url}
\usepackage{xparse}
\usepackage{makecell}
\usepackage{varwidth}
\usepackage{bm}
\usepackage{arydshln}
\usepackage{comment}
\usepackage{wrapfig}

\usepackage{caption}
\usepackage{float}
\usepackage{booktabs}

\usepackage{nidanfloat}
\usepackage{subcaption}

\usepackage{mathtools}

\usepackage{booktabs}

\usepackage{float}

\usepackage{multirow}

\newcolumntype{C}[1]{>{\centering\let\newline\\\arraybackslash\hspace{0pt}}m{#1}}

\usepackage[normalem]{ulem}

\usepackage{amsmath,pgfplots,amssymb}
\usepackage{subcaption,graphicx}

\newtheorem{theorem}{Theorem}

\newtheorem{lemma}{Lemma}

\theoremstyle{definition}

\newtheorem{definition}{Definition}

\theoremstyle{definition}
\newtheorem{remark}{Remark}

\theoremstyle{definition}

\newcommand{\F}{\mathbb{F}}

\DeclareMathOperator{\terms}{terms}

\usetikzlibrary{matrix,decorations.pathreplacing}

\def \black{\color{black}}

\newcommand{\update}[1]{\textcolor{black}{#1}}

\ifCLASSINFOpdf
\else
\fi
\begin{document}
%
\title{GASP Codes for Secure Distributed Matrix Multiplication}
%
%
%

\author{Rafael G. L. D'Oliveira, Salim El Rouayheb, David Karpuk \thanks{Authors listed in alphabetical order of last name. The first two authors were partially supported by the NSF under Grant CNS-1801630.   Part of this work was completed while D.\ Karpuk was visiting the research group of S.\ El Rouayheb at Rutgers University.  \update{R.G.L. D'Oliveira is currently with the Massachusetts Institute of Technology. D.\ Karpuk is currently with F-Secure Corporation, Helsinki, Finland.}}  \\ ECE, Rutgers University, Piscataway, NJ\\ Departamento de Matem\'aticas, Universidad de los Andes, Bogot\'a, Colombia \\ Emails: rafaeld@mit.edu, salim.elrouayheb@rutgers.edu, davekarpuk@gmail.com }

\maketitle


\begin{abstract}

We consider the problem of secure distributed matrix multiplication (SDMM) in which a user wishes to compute the product of two matrices with the assistance of honest but curious servers. We construct polynomial codes for SDMM  by  studying a combinatorial problem on a special type of addition table, which we call the degree table. The codes are based on  arithmetic progressions, and are thus named GASP (Gap Additive Secure Polynomial) Codes.  GASP Codes are shown to outperform all previously known polynomial codes for secure distributed matrix multiplication in terms of download rate.

\end{abstract}

%
\IEEEpeerreviewmaketitle

\section{Introduction}




 We consider the problem of secure distributed matrix multiplication (SDMM), in which a user has two matrices $A$ and $B$ and wishes to compute their product $AB$ with the assistance of $N$ servers, without leaking any information about $A$ or $B$ to any server.  We assume that all servers are honest and responsive, but that they are curious, in that any $T$ of them may collude to try to deduce information about either $A$ or $B$.  

When considering the problem of SDMM from an information-theoretic perspective, the primary performance metric used in the literature is that of the \emph{download rate}, or simply \emph{rate}, which we denote by $\mathcal{R}$.  \update{In our scenario, the user queries the servers to perform various matrix mulitplications, and the servers respond with answers that the user can use to piece together the final desired result $AB$.  In this admittedly heuristic description, the rate $\mathcal{R}$ is the ratio of the size of the desired result $AB$ (in bits) to the total amount of information (in bits) the user downloads to obtain the answers from the servers.}  The goal is to construct a SDMM scheme with rate $\mathcal{R}$ as large as possible. 

The problem of constructing polynomial codes for SDMM can be  summarized as follows. We partition the matrices $A$ and $B$ as follows:
\begin{equation}\label{partition1}
    A = \begin{bmatrix}
    A_1 \\ \vdots \\ A_K
    \end{bmatrix},\quad 
    B = \begin{bmatrix} 
    B_1 & \cdots & B_L
    \end{bmatrix},\quad \text{so that}\quad 
    AB = \begin{bmatrix}
    A_1B_1 & \cdots & A_1B_L \\
    \vdots & \ddots & \vdots \\
    A_KB_1 & \cdots & A_KB_L
    \end{bmatrix},
\end{equation}
making sure that all products $A_kB_\ell$ are well-defined and of the same size.  Clearly, computing the product $AB$ is equivalent to computing all subproducts $A_kB_\ell$.  One then constructs a polynomial $h(x)$ whose coefficients encode the submatrices $A_kB_\ell$, and has $N$ servers compute evaluations $h(a_1),\ldots,h(a_N)$.  The polynomial $h$ is constructed so that every $T$-subset of evaluations reveals no information about $A$ or $B$, but so that the user can reconstruct all of $AB$ given all $N$ evaluations. This follows the general mantra of evaluation codes and, in particular, polynomial codes as originally introduced in \cite{polycodes1} and \cite{polycodes2}.  

One can view the parameters $K$ and $L$ as controlling the complexity of the matrix multiplication operations the servers must perform. Imagine a scenario in which one may hire as many servers $N$ as one wants to assist in the SDMM computation, but the computational capacity of each server is limited.  In this scenario, one may have fixed values of $K$ and $L$, and then maximizing the rate $\mathcal{R}$ becomes a question of minimizing $N$.  This is the general perspective we adopt in the SDMM problem.

\subsection{Related Work} 

Let $A$ and $B$ be partitioned as in (\ref{partition1}), and consider the problem of SDMM with $N$ servers and $T$-security.  In \cite{ravi2018mmult}, a distributed matrix multiplication scheme is presented for the case $K = L$ which achieves a download rate of 
\begin{equation}
    \mathcal{R}_1 = \frac{K^2}{(K+T)^2}, \quad \text{or equivalently}\quad \mathcal{R}_1 = \frac{\left(\left\lceil\sqrt{N} - T \right\rceil\right)^2}{\left(\left\lceil \sqrt{N} - T\right\rceil + T\right)^2}.
\end{equation}
In \cite{kakar}, this is improved to 
\begin{equation} \label{eq:kakar rate}
    \mathcal{R}_2 = \frac{KL}{ (K+T)(L+1)-1 },
\end{equation}
where the polynomial code uses $N = (K+T)(L+1)-1$ servers.  Given some fixed $N$ and $T$, the authors of \cite{kakar} then find a near-optimal solution to the problem of finding $K$ and $L$ such that $(K+T)(L+1)-1\leq N$ and that the rate $\mathcal{R}_2$ as above is maximized.  In \cite{koreans}, the authors study the case of $T = 1$ and obtain a download rate of $\mathcal{R} = KL/(KL+K+L)$, which is the rate of \cite{kakar} in this case.  As far as the present authors are aware, \cite{ravi2018mmult,kakar,koreans} are the only works currently in the literature which study SDMM from the information-theoretic perspective.

We distinguish the SDMM problem from the case where only one of the matrices must be kept secure. In this case, one can use methods like Shamir's secret sharing \cite{sham79} or Staircase codes \cite{bita17}, if one is also interested in straggler mitigation. 

Polynomial codes were originally introduced in \cite{polycodes1} in a slightly different setting, namely to mitigate stragglers in distributed matrix multiplication. This work was followed up by \cite{polycodes2} which studied fundamental limits of this problem, introduced a generalization of polynomial codes known as \emph{entangled} polynomial codes, and applied similar ideas to other problems in distributed computing.  In \cite{pulkit}, the authors develop MatDot and PolyDot codes for distributed matrix multiplication with stragglers, and show that while the communication cost is higher than that of the polynomial codes of \cite{polycodes1}, the recovery threshold, defined to be the minimum number of workers which need to respond to guarantee successful decoding, is much smaller than that of \cite{polycodes1}.  The MatDot codes of \cite{pulkit} were then applied to the problem of nearest neighbor estimation in \cite{pulkit2}.  More fundamental questions about the trade-off in computation cost and communication cost in distributed computing were previously addressed in \cite{fundamental}. \update{However, the polynomial codes in these aforementioned works are not designed to ensure security, making them not applicable to settings where there are privacy concerns related to the data being used. This type of setting could range from training neural networks on personal devices to computations on medical data, where legislation requires that certain privacy conditions are met.}

\update{Another line of work is Lagrange Coded Computing, a polynomial coding strategy introduced in \cite{lagrange} to mitigate stragglers and adversaries in distributed polynomial coded computation.  The results in \cite{lagrange} focus on minimizing the number of required servers for the computation subject to privacy, robustness, and polynomial degree constraints.  However, applying the ideas of \cite{lagrange} to the current scenario yields only one-sided privacy, wherein either $A$ or $B$ is kept private, but not both.  More related to the current work is that of Private Polynomial Computation \cite{ppc}, which does provide two-sided privacy, but focuses on generic strategies which work for all polynomials of a given degree, rather than polynomial coding strategies tailored for the problem of matrix multiplication.  Lastly, it seems that concerns related to data partitioning and block length make the results of the present paper (and generally results on using polynomnial codes for SDMM) incomparable with those of \cite{lagrange} or \cite{ppc}.}

\subsection{Main Contribution}

The main contributions of this work are as follows.
\begin{itemize}

    \item In Section \ref{sec:MainExample} we introduce our polynomial code $\mathsf{GASP}$ via an explicit example, in order to demonstrate all of the subtleties of the scheme construction.  In Section \ref{sec:polycodes} we formalize the notion of a polynomial code and introduce basic definitions.
    
    \item In Section \ref{sec:combinatorics} we introduce the key notion of this paper, the \emph{degree table} of a polynomial code. We prove, in Theorem~\ref{blackbox}, that to every degree table corresponds a secure distributed matrix multiplication scheme.
    
    \item In Section \ref{sec:gaspbig}, we present a secure distributed matrix multiplication scheme, $\mathsf{GASP}_{\text{big}}$. We show that $\mathsf{GASP}_{\text{big}}$ outperforms, for almost all parameters, all previously known schemes in the literature, in terms of the download rate.
    
    \item In Section \ref{sec:gaspsmall}, we present a secure distributed matrix multiplication scheme, $\mathsf{GASP}_{\text{small}}$. We show that $\mathsf{GASP}_{\text{small}}$ outperforms $\mathsf{GASP}_{\text{big}}$ when $T < \min \{K,L\}$.
    
    \item In Section \ref{sec:combining}, we present a secure distributed matrix multiplication scheme, $\mathsf{GASP}$, by combining both $\mathsf{GASP}_{\text{small}}$ and $\mathsf{GASP}_{\text{big}}$. $\mathsf{GASP}$ outperforms all previously known schemes, for almost all parameters, in terms of the download rate.
    
    The rate of $\mathsf{GASP}$, for $L\leq K$,  is given in Table \ref{tab:gasp rate}. For $K<L$, the rate is given by interchanging $K$ and $L$.
    
\begin{table}[h] 
    \centering
    \begin{tabular}{c c}
       Download Rate  & Regions \\
       \toprule
        $\dfrac{KL}{KL+K+L}$ & $1 = T < L \leq K$  \\[13pt]
        $\dfrac{KL}{KL+K+L+T^2+T-3}$ & $2\leq T < L \leq K$ \\[13pt]
        $\dfrac{KL}{(K+T)(L+1) - 1}$ & $ L \leq T < K$ \\[13pt]
        $\dfrac{KL}{2KL + 2T - 1}$ & $L \leq K \leq T$
    \end{tabular}
    \caption{The download rate of GASP codes.}
    \label{tab:gasp rate}
\end{table}


\end{itemize}

\begin{figure}[H]
    \centering
    \includegraphics[width=.3\textwidth]{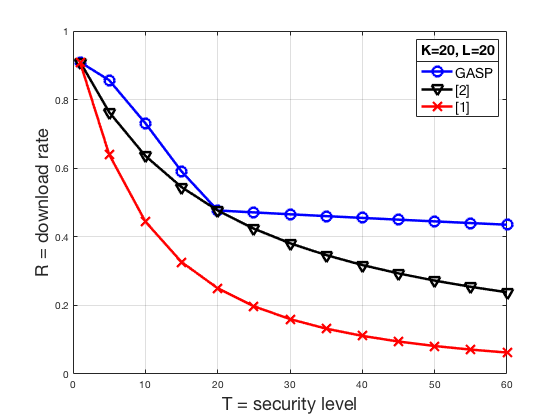} \hspace{2.5em}
    \includegraphics[width=.3\textwidth]{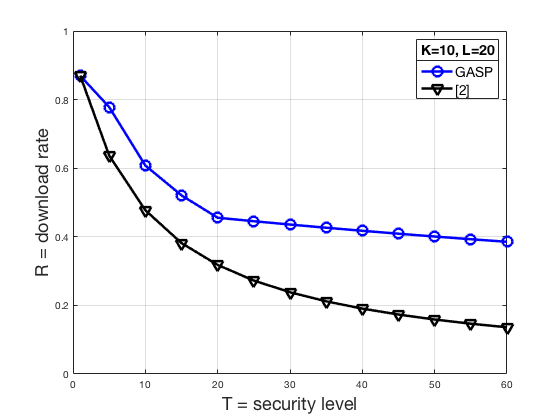} \\
    \caption{Comparison of the Polynomial Code $\mathsf{GASP}$ with that of \cite{ravi2018mmult} and \cite{kakar}. We plot the rate of the schemes for $K = 20$ and $L = 20$ on the left, and $K = 10$ and $L = 20$ on the right.}
    \label{fig:compare1}
\end{figure}

\update{We plot in Fig.\ \ref{fig:compare1} the rates obtained by GASP against those obtained by the polynomial code SDMM strategies of \cite{ravi2018mmult,kakar}.  Before launching into the construction of GASP, let us offer some intuitive explanation as to the large improvement in rate offered by GASP over \cite{ravi2018mmult,kakar}.  The polynomial codes of \cite{ravi2018mmult,kakar}, as well as those of the current work, all have the user decode the necessary blocks of $AB$ by interpolating a polynomial $h(x)$, and obtaining the $A_kB_{\ell}$ as coefficients of this polynomial.  The rate of all three strategies is completely determined by how many evaluations $h(x)$ requires to be interpolated completely, as this is the number of servers employed by the user.  The strategies of \cite{ravi2018mmult,kakar} force every coefficient of $h(x)$ to be potentially non-zero, and therefore interpolating $h(x)$ requires $\deg(h(x))+1$ evaluations.  In contrast, GASP codes purposefully rig up $h(x)$ so that it has as many zero coefficients as possible, and that the user knows where these zero coefficients are located.  This allows the user to interpolate $h(x)$ with substantially fewer than the expected number $\deg(h(x)) + 1$ of evaluations.  While the polynomials $h(x)$ from the current work and those of \cite{ravi2018mmult,kakar} have different degrees for the same parameters $K$, $L$, and $T$, this extra flexibility still allows us to generally use substantially fewer servers than the polynomial codes of \cite{ravi2018mmult,kakar}.}

\section{A Motivating Example: $K = L = 3$ and $T = 2$}\label{sec:MainExample}

We begin our scheme description with the following example, which we present in as much detail as possible to showcase the essential ingredients of the scheme.  In this example \update{a user wishes to multiply two matrices $A$ and $B$ over a finite field $\F_q$, which are selected independently and uniformly at random from their respective ambient spaces. The user partitions the matrices} as:
\[
A = \begin{bmatrix}
A_1 \\ A_2 \\ A_3
\end{bmatrix},\quad
B = \begin{bmatrix}
B_1 & B_2 & B_3
\end{bmatrix}
\]
so that all products $A_kB_\ell$ are well-defined and of the same size.  The product $AB$ is given by
\[
AB = \begin{bmatrix}
A_1B_1 & A_1B_2 & A_1B_3 \\
A_2B_1 & A_2B_2 & A_2B_3 \\
A_3B_1 & A_3B_2 & A_3B_3
\end{bmatrix}
\]
We construct a scheme which computes each term $A_kB_\ell$, and therefore all of $AB$, via polynomial interpolation. The scheme is private against any $T=2$ servers colluding to deduce the identities of $A$ and $B$, and uses a total of $N = 18$ servers.

Let $R_1$ and $R_2$ be two matrices picked independently and uniformly at random with entries in $\F_q$, both of size equal to the $A_k$.  Similarly, pick $S_1$ and $S_2$ independently and uniformly at random of size equal to that of the $B_\ell$.  Define polynomials
\[
\begin{aligned}
f(x) &= A_1x^{\alpha_1} + A_2x^{\alpha_2} + A_3x^{\alpha_3} + R_1x^{\alpha_4} + R_2x^{\alpha_5} \\
g(x) &= B_1x^{\beta_1} + B_2x^{\beta_2} + B_3x^{\beta_3} + S_1x^{\beta_4} + S_2x^{\beta_5} \\
\end{aligned}
\]
where the $\alpha_k$ and $\beta_\ell$ are natural numbers that will be determined shortly.  

As in \cite{ravi2018mmult}, we will recover the products $A_kB_\ell$ by interpolating the product $h(x) = f(x)g(x)$.  Specifically, for some evaluation points $a_n\in \F_q$, we will send $f(a_n)$ and $g(a_n)$ to server $n = 1,\ldots,N$, who then responds with $h(a_n) = f(a_n)g(a_n)$. These evaluations will suffice to interpolate all of $h(x)$.  In particular, we will be able to retrieve the coefficients of $h(x)$, which in turn will allow us to decode all the $A_kB_\ell$.

The product $h(x) = f(x)g(x)$ is given by
\[
h(x) = \sum_{1\leq k,\ell\leq 3}A_kB_\ell x^{\alpha_k+\beta_\ell} + \sum_{\substack{1\leq k\leq 3 \\ 4\leq \ell \leq 5}}A_kS_\ell x^{\alpha_k+\beta_\ell} + \sum_{\substack{4\leq k\leq 5 \\ 1\leq \ell \leq 3}}B_\ell R_k x^{\alpha_k+\beta_\ell} + \sum_{4\leq k,\ell\leq 5} R_k S_\ell x^{\alpha_k+\beta_\ell}
\]
We wish to assign the exponents $\alpha_k$ and $\beta_\ell$ to guarantee decodability.  Consider the following condition on the exponents:
\[\alpha_k + \beta_\ell \neq \alpha_{k'} + \beta_{\ell'}\quad \text{for all $(k,\ell)\in [3]\times [3]$ and all $(k',\ell')\in[5]\times[5]$ such that $(k,\ell) \neq (k',\ell')$.}
\]
That is, all of the exponents corresponding to the terms we wish to decode must be distinct from all the other exponents appearing in $h(x)$.  This guarantees that each product $A_kB_\ell$ appears as the unique coefficient of a unique power of $x$.  The immediate goal is to minimize the number of distinct powers of $x$ appearing in $h(x)$, subject to the above condition.  This will allow us to minimize the number of servers used by the scheme, thereby maximizing the rate.

The problem of assigning the $\alpha_k$ and $\beta_\ell$ can alternately be phrased as the following combinatorial problem.  Consider the following addition table:

\begin{table}[!htb]
\centering
\begin{tabular}{c|ccc|cc}
 & $\beta_1$ & $\beta_2$ & $\beta_3$ & $\beta_4$ & $\beta_5$ \\ 
\toprule
$\alpha_1$ & $\alpha_1 + \beta_1$ & $\alpha_1 + \beta_2$ & $\alpha_1 + \beta_3$ & $\alpha_1$ + $\beta_4$ & $\alpha_1$ + $\beta_5$ \\
$\alpha_2$ & $\alpha_2 + \beta_1$ & $\alpha_2 + \beta_2$ & $\alpha_2 + \beta_3$ & $\alpha_2$ + $\beta_4$ & $\alpha_2$ + $\beta_5$ \\
$\alpha_3$ & $\alpha_3 + \beta_1$ & $\alpha_3 + \beta_2$ & $\alpha_3 + \beta_3$ & $\alpha_3$ + $\beta_4$ & $\alpha_3$ + $\beta_5$ \\
\midrule
$\alpha_4$ & $\alpha_4 + \beta_1$ & $\alpha_4 + \beta_2$ & $\alpha_4 + \beta_3$ & $\alpha_4$ + $\beta_4$ & $\alpha_4$ + $\beta_5$ \\
$\alpha_5$ & $\alpha_5 + \beta_1$ & $\alpha_5 + \beta_2$ & $\alpha_5 + \beta_3$ & $\alpha_5$ + $\beta_4$ & $\alpha_5$ + $\beta_5$ \\
\bottomrule
\end{tabular}
\end{table}
\noindent We call this table the \emph{degree table} since it encodes the degrees that appear in $h(x) = f(x)g(x)$. With this in mind, we wish to pick $\alpha_k,\beta_\ell\in\mathbb{N}$ such that every term in the upper-left $3\times3$ block is distinct from every other number in the table.  Outside this block, we wish to minimize the number of distinct integers that appear, in order to minimize the number of non-zero coefficients of $h(x)$ and therefore the number of required evaluation points.

Consider the assignment
\[
\alpha_1 = 0,\ \alpha_2 = 1,\ \alpha_3 = 2,\ \alpha_4 = 9,\ \alpha_5 = 12 \quad\text{and}\quad
\beta_1 = 0,\ \beta_2 = 3,\ \beta_3 = 6,\ \beta_4 = 9,\ \beta_5 = 10,
\]
for which the degree table becomes

\begin{table}[H]
\centering
\begin{tabular}{c|ccc|cc}

 & $\beta_1 = 0$ & $\beta_2 = 3$ & $\beta_3 = 6$ & $\beta_4 = 9$ & $\beta_5 = 10$ \\ 
\toprule
$\alpha_1 = 0$ & $0$ & $3$ & $6$ & $9$ & $10$ \\
$\alpha_2 = 1$ & $1$ & $4$ & $7$ & $10$ & $11$ \\
$\alpha_3 = 2$ & $2$ & $5$ & $8$ & $11$ & $12$ \\
\midrule
$\alpha_4 = 9$ & $9$ & $12$ & $15$ & $18$ & $19$ \\
$\alpha_5 = 12$ & $12$ & $15$ & $18$ & $21$ & $22$ \\
\bottomrule
\end{tabular}
\end{table}
\noindent which satisfies our decodability condition.  Concretely, the polynomial $h(x)$ is now of the form
\[
\begin{aligned}
h(x) = A_1B_1 + \cdots + A_3B_3 x^8 + C_9x^9 + C_{10}x^{10} + C_{11}x^{11} + C_{12}x^{12} + C_{15}x^{15} + C_{18}x^{18} + C_{19}x^{19} + C_{21}x^{21} + C_{22}x^{22}
\end{aligned}
\]
which has $N = 18$ potentially non-zero coefficients.  Here each $C_j$ is a sum of products of matrices where each summand has either $R_k$ or $S_\ell$ as a factor, and thus their precise nature is not important for decoding.  We now show that over a suitable field $\F_q$, we can find $N  = 18$ evaluation points $a_n$ which suffice to interpolate $h(x)$, \update{even though $\deg(h(x)) = 22$.  This difference is subtle but crucial: the user knows exactly which coefficients of $h(x)$ are zero, and can thus interpolate the entire polynomial with fewer than the $\deg(h(x))+1$ evaluations one would normally need.  This is in stark contrast with the strategies of \cite{ravi2018mmult,kakar}, where $f(x)$ and $g(x)$ are constructed so that every coefficient of $h(x)$ is non-zero (though for the same parameters $K$, $L$, and $T$, the polynomials $h(x)$ from \cite{ravi2018mmult,kakar} are of a different degree than the $h(x)$ we obtain).}

Let $\mathcal{J}$ be the set of exponents which occur in the above expression for $h(x)$, that is,
\[
\mathcal{J} = \{0,\ldots,8,9,10,11,12,15,18,19,21,22\}
\]
so that $|\mathcal{J}| = 18$.  We wish to find an evaluation vector ${\bf a} = (a_1,\ldots,a_N)\in \F_q^N$ such that the $18\times 18$ generalized Vandermonde matrix
\[
GV({\bf a},\mathcal{J}) = \begin{bmatrix}
a_n^j
\end{bmatrix},\quad 1\leq n\leq 18,\ j\in \mathcal{J}
\]
is invertible. One can easily check that for $q = 29$, the assignment $a_n = n \pmod{29}$ for $n = 1,\ldots,18$ results in ${\det(GV({\bf a},\mathcal{J})) = 20\neq 0\pmod{29}}$.  Thus the coefficients of $h(x)$, in particular the $A_kB_\ell$, are uniquely decodable in the current scheme.

It is perhaps not obvious that the scheme we have described satisfies the $2$-privacy condition.  Let us show that this is indeed the case.  \update{As in Example 1} in \cite{ravi2018mmult}, the $2$-privacy condition will be satisfied provided that the matrices
\[
P_{n,m} = \begin{bmatrix}
a_n^{9} & a_n^{12} \\
a_m^{9} & a_m^{12}
\end{bmatrix},\quad
Q_{n,m} = \begin{bmatrix}
a_n^{9} & a_n^{10} \\
a_m^{9} & a_m^{10}
\end{bmatrix}
\]
are invertible for any pair $1\leq n\neq m\leq 18$.  We compute
\[
\det(Q_{n,m}) = a_n^9a_m^9(a_m-a_n)\quad\text{and}\quad\det(P_{n,m}) = a_n^9a_m^9(a_m^3-a_n^3)
\]
thus provided that $a_n^3\neq a_m^3$ for all $n\neq m$, and none of the $a_n$ are zero, these matrices will all be invertible.  However, we have $\gcd(3,q-1) = 1$, thus the map $x\mapsto x^3$ is a bijection from $\F_{29}$ to itself.  Thus $a_n\neq a_m$ implies that $a_n^3\neq a_m^3$ for all $n\neq m$, and we see that the determinants of the above matrices are all non-zero and hence the $2$-privacy condition is satisfied.

For $K = L = 3$ and $T = 2$, let $\mathcal{R}_1$ denote the download rate of \cite{kakar} and $\mathcal{R}_2$ that of \cite{ravi2018mmult}. We have
\[
\mathcal{R}_1 = \frac{K^2}{(K+T)(K+1)-1} = \frac{9}{19}\quad\text{and}\quad\mathcal{R}_2 = \frac{K^2}{(K+T)^2} = \frac{9}{25}
\]
whereas the scheme we have presented above improves on these constructions to achieve a rate of
\[
\mathcal{R} = \frac{K^2}{K^2 + 2K + T^2 + T - 3} = \frac{9}{18} = \frac{1}{2}.
\]
While this improvement in this example is marginal, we will see later that for large parameters we achieve significant gains over the polynomial codes of \cite{ravi2018mmult,kakar}.

\section{Polynomial Codes}\label{sec:polycodes}

Let $A$ and $B$ be matrices over a finite field $\F_q$, \update{selected by a user independently and uniformly at random from the set of all matrices of their respective sizes, and} partitioned as in equation (\ref{partition1}) so that all products $A_kB_\ell$ are well-defined and of the same size.   Then $AB$ is the block matrix $AB = \left(A_kB_\ell\right)_{1\leq k\leq K, 1\leq \ell\leq L}$. A \emph{polynomial code} is a tool for computing the product $AB$ in a distributed manner, by computing each block $A_kB_\ell$.  Formally, we define a polynomial code as follows.

\begin{definition}
The \emph{polynomial code} $\mathsf{PC}(K,L,T,N,\alpha,\beta)$ consists of the following data:
\begin{itemize}
    \item[(i)] positive integers $K$, $L$, $T$, and $N$,
    \item[(ii)] $\alpha = (\alpha_1,\ldots,\alpha_{K+T})\in\mathbb{N}^{K+T}$, and
    \item[(iii)] $\beta = (\beta_1,\ldots,\beta_{L+T})\in\mathbb{N}^{L+T}$.
\end{itemize}
\end{definition}

A polynomial code $\mathsf{PC}(K,L,T,N,\alpha,\beta)$ is used to securely compute the product $AB$ as follows.  A user chooses $T$ matrices $R_t$ over $\F_q$ of the same size as the $A_k$ independently and uniformly at random, and $T$ matrices $S_t$ of the same size as the $B_\ell$ independently and uniformly at random.  They define polynomials $f(x)$ and $g(x)$ by
\[
f(x) = \sum_{k = 1}^KA_kx^{\alpha_k} + \sum_{t = 1}^TR_tx^{\alpha_{K+t}}\quad\text{and}\quad
g(x) = \sum_{\ell = 1}^LB_\ell x^{\beta_\ell} + \sum_{t = 1}^TS_t x^{\beta_{L+t}}
\]
and let
\begin{equation}\label{productpoly}
h(x) = f(x)g(x).
\end{equation}
Given $N$ servers, a user chooses evaluation points $a_1,\ldots,a_N\in \F_{q^r}$ in some finite extension $\F_{q^r}$ of $\F_q$.  They then send $f(a_n)$ and $g(a_n)$ to server $n = 1,\ldots,N$, who computes the product $f(a_n)g(a_n) = h(a_n)$ and transmits it back to the user.  The user then interpolates the polynomial $h(x)$ given all of the evaluations $h(a_n)$, and attempts to recover all products $A_kB_\ell$ from the coefficients of $h(x)$.  We omit the evaluation vector ${\bf a}$ from the notation $\mathsf{PC}(K,L,T,N,\alpha,\beta)$ because as we will shortly show, it does not really affect any important analysis of the polynomial code.

\begin{definition}\label{pcdecodablesecure}
A polynomial code $\mathsf{PC}(K,L,T,N,\alpha,\beta)$ is \emph{decodable and $T$-secure} if there exists some evaluation vector ${\bf a} = (a_1,\ldots,a_N)\in\F_{q^r}^N$ for some $r > 0$ such that for any $A$ and $B$ as above, the following two conditions hold.
\begin{itemize}
    \item[(i)] (Decodability) All products $A_kB_\ell$ for $k = 1,\ldots,K$ and $\ell = 1,\ldots,L$ are completely determined by the evaluations $h(a_n)$ for $n = 1,\ldots,N$.
    \item[(ii)] ($T$-security) For any $T$-tuple $\{n_1,\ldots,n_T\}\subseteq[N]$, we have
\[
I(f(a_{n_1}),g(a_{n_1}),\ldots,f(a_{n_T}),g(a_{n_T});A,B) = 0.
\]
where $I(\cdot;\cdot)$ denotes mutual information between two random variables.
\end{itemize}
\end{definition}

\begin{definition}
Suppose that the polynomial code $\mathsf{PC}(K,L,T,N,\alpha,\beta)$ is decodable and $T$-secure.  The \emph{download rate}, or simply the \emph{rate}, of this polynomial code is defined to be
\[
\mathcal{R} = \frac{KL}{N}.
\]
\end{definition}

Given parameters $K$, $L$, and $T$, the goal of polynomial coding is to construct a decodable and $T$-secure polynomial code $\mathsf{PC}(K,L,T,\alpha,\beta)$ with download rate as large as possible.  This is equivalent to minimizing the number of servers $N$, or equivalently, the number of evaluation points needed by the code.

\section{The Degree Table}\label{sec:combinatorics}


In this section we relate the construction of polynomial codes for SDMM with a certain combinatorial problem.  This connection will guide our constructions and aid us in proving that our polynomial codes are decodable and $T$-secure.

\begin{definition}
Let $\alpha \in \mathbb{N}^{K+T}$ and $\beta \in \mathbb{N}^{L+T}$. The \emph{outer sum} $\alpha\oplus\beta\in \mathbb{N}^{(K+T)\times (L+T)}$ of $\alpha$ and $\beta$ is defined to be the matrix
\[
\alpha \oplus\beta = 
\begin{bmatrix}
\alpha_1+\beta_1 & \cdots & \alpha_1 + \beta_{L+T} \\ 
\vdots & \ddots & \vdots \\ 
\alpha_{K+T} + \beta_1 & \cdots & \alpha_{K+T} + \beta_{L+T}
\end{bmatrix}.
\]
\end{definition}





\begin{definition}\label{outersumsecure}
Let $\alpha \in \mathbb{N}^{K+T}$ and $\beta \in \mathbb{N}^{L+T}$. We say that the outer sum $\alpha \oplus \beta$ is \emph{decodable and $T$-secure} if the following two conditions hold:
\begin{itemize}
    \item[(i)] (Decodability) $(\alpha\oplus\beta)_{k,\ell}\neq (\alpha\oplus\beta)_{k',\ell'}$ for all $(k,\ell)\in [K]\times [L]$ and all $(k',\ell')\in [K+T]\times [L+T]$.
    \item[(ii)] ($T$-security) $\alpha_{K+t} \neq \alpha_{K+t'}$ and $\beta_{L+t} \neq \beta_{L+t'}$ for every $t\neq t'\in[T]$.
\end{itemize}
\end{definition}

Constructing $\alpha$ and $\beta$ so that $\alpha\oplus\beta$ is decodable and $T$-secure can be realized as the following combinatorial problem, displayed in Table \ref{problem}.  The condition of decodability from Definition \ref{outersumsecure} simply states that each $\alpha_k+\beta_\ell$ in the red block must be distinct from every other entry in $\alpha\oplus\beta$.  The condition of $T$-security states that all $\alpha_{K+t}$ in the green block must be pairwise distinct, and all $\beta_{L+t}$ in the blue block must be pairwise distinct. We refer to this table as the degree table.

\begin{table}[!htb]
\centering
\resizebox{.8\textwidth}{!}{
\begin{tabular}{c|ccc|ccc}

 & $\beta_1$ & $\cdots$ & $\beta_L$ &  \cellcolor{blue!25} $\beta_{L+1}$ & \cellcolor{blue!25} $\cdots$ & \cellcolor{blue!25} $\beta_{L+T}$  \\ 
\toprule
$\alpha_1$ & \cellcolor{red!25} $\alpha_1 + \beta_1$ & \cellcolor{red!25} $\cdots$ & \cellcolor{red!25} $\alpha_1 + \beta_L$ & $\alpha_1 + \beta_{L+1}$ & $\cdots$ & $\alpha_1 + \beta_{L+T}$  \\
$\vdots$ & \cellcolor{red!25} $\vdots$ & \cellcolor{red!25} $\ddots$ & \cellcolor{red!25} $\vdots$ & $\vdots$ & $\ddots$ & $\vdots$  \\
$\alpha_K$ & \cellcolor{red!25} $\alpha_K + \beta_1$ & \cellcolor{red!25} $\cdots$ & \cellcolor{red!25} $\alpha_K + \beta_L$ & $\alpha_K + \beta_{L+1}$ & $\cdots$ & $\alpha_K + \beta_{L+T}$  \\
\midrule
\cellcolor{green!25} $\alpha_{K+1}$ &  $\alpha_{K+1} + \beta_1$ & $\cdots$ &  $\alpha_{K+1} + \beta_L$ & $\alpha_{K+1} + \beta_{L+1}$ & $\cdots$ & $\alpha_{K+1} + \beta_{L+T}$ \\
\cellcolor{green!25} $\vdots$ &  $\vdots$ & $\ddots$ &  $\vdots$ & $\vdots$ & $\ddots$ & $\vdots$ \\
\cellcolor{green!25} $\alpha_{K+T}$ &  $\alpha_{K+T} + \beta_1$ & $\cdots$ &  $\alpha_{K+T} + \beta_L$ & $\alpha_{K+T} + \beta_{L+1}$ & $\cdots$ & $\alpha_{K+T} + \beta_{L+T}$ \\
\bottomrule
\end{tabular}
}\caption{The combinatorial problem of constructing $\alpha$ and $\beta$ so that $\alpha\oplus\beta$ is decodable and $T$-secure.}\label{problem}
\end{table}


\begin{definition}
Let $A$ be a matrix with entries in $\mathbb{N}$. We define the \emph{terms} of $A$ to be the set
\[ \terms A = \{ n \in \mathbb{N}: \exists (i,j), A_{ij}=n \}. \]
\end{definition}

The next lemma and theorem allow us to reduce the construction of Polynomial Codes for SDMM to the combinatorial problem of constructing $\alpha$ and $\beta$ such that the degree table, $\alpha\oplus\beta$, is decodable and $T$-secure.  The proof of the lemma is straightforward and thus omitted.

\begin{lemma}\label{termsterms}
Consider the polynomial code $\mathsf{PC}(K,L,T,N,\alpha,\beta)$, with associated polynomials
\[
f(x) = \sum_{k = 1}^KA_kx^{\alpha_k}\quad\text{and}\quad g(x) = \sum_{\ell = 1}^LB_\ell x^{\beta_\ell}.
\]
 Then we can express the product $h(x)$ of $f(x)$ and $g(x)$ as
\begin{equation}\label{hterms}
h(x) = f(x)g(x) = \sum_{j\in \mathcal{J}}C_jx^j
\end{equation}
for some matrices $C_j$, where $\mathcal{J} = \terms(\alpha\oplus\beta)$.
\end{lemma}

Thus, the terms in the outer sum $\alpha \oplus \beta$ correspond to the terms in the polynomial $h(x)=f(x)g(x)$. Because of this, we refer to the table representation of $\alpha \oplus \beta$ in Table \ref{problem} as the \emph{degree table} of the polynomial code $\mathsf{PC}(K,L,T,N,\alpha,\beta)$.  The following theorem allows us to reduce the construction of polynomial codes to the construction of degree tables which are decodable and $T$-secure.

\begin{theorem}\label{blackbox}
Let $\mathsf{PC}(K,L,T,N,\alpha,\beta)$ be a polynomial code, where $N = |\terms(\alpha\oplus\beta)|$.  Suppose that the degree table, $\alpha\oplus\beta$, satisfies the decodability and $T$-security conditions of Definition \ref{outersumsecure}.  Then the polynomial code $\mathsf{PC}(K,L,T,N,\alpha,\beta)$ is decodable and $T$-secure.
\end{theorem}

\begin{IEEEproof}
The proof is an application of the Schwarz-Zippel Lemma.  One finds sufficient conditions for decodability and $T$-security that reduce to the simultaneous non-vanishing of $2\binom{N}{T}+1$ determinants.  One can find a point ${\bf a}\in \F_{q^r}^N$ for some $r>0$ at which none of these polynomials is zero.  We relegate a detailed proof to the Appendix.
\end{IEEEproof}

Thanks to Theorem \ref{blackbox}, constructing a polynomial code scheme for secure distributed matrix multiplication can be done by constructing $\alpha$ and $\beta$ such that the degree table, $\alpha\oplus\beta$, is decodable and $T$-secure.  For this reason, the visualization in Table~\ref{problem} is extremely useful, both as a guide for constructing polynomial codes for SDMM and as a method for calculating the corresponding download rate.  In this context, maximizing the download rate is equivalent to minimizing $|\terms(\alpha\oplus\beta)|$, the number of distinct integers in the degree table shown in Table \ref{problem}, subject to decodability and $T$-security.

\begin{remark}
Suppose that the degree table, $\alpha\oplus\beta$, is decodable and $T$-secure, and let $\mathbb{K}$ be an algebraic closure of $\F_q$.  One can show that the set of all ${\bf a}= (a_1,\ldots,a_N)$ such that $\mathsf{PC}(K,L,T,N,\alpha,\beta)$ is decodable and $T$-secure is a Zariski open subset of $\mathbb{K}^N$.  In practice, this means that given $K,L,T,\alpha,\beta$, if we choose ${\bf a}\in\F_{q^r}^N$ uniformly at random, then the probability that the polynomial code $\mathsf{PC}(K,L,T,N,\alpha,\beta)$ is decodable and $T$-secure goes to $1$ as $r\rightarrow\infty$.  Thus, finding such evaluation vectors is not a difficult task.
\end{remark}

\section{A Polynomial Code for Big $T$} \label{sec:gaspbig}

In this section, we construct a polynomial code, $\mathsf{GASP}_{\text{big}}$, which has better rate than all previous schemes in the literature. The scheme construction chooses $\alpha$ and $\beta$ to attempt to minimize the number of distinct integers in the degree table, $\alpha\oplus\beta$. The scheme construction proceeds by choosing $\alpha_k$ and $\beta_\ell$ to belong to certain arithmetic progressions, and minimizes the number of terms in the lower-right $T\times T$ block of the degree table, $\alpha\oplus\beta$, shown in Table \ref{bigmatrix2}.

\begin{definition} \label{def:gaspbig}
Given $K$, $L$, and $T$, define the polynomial code $\mathsf{GASP}_{\text{big}}$ as follows.  Let $\alpha$ and $\beta$ be given by
\begin{equation}\label{alphabetadefn2}
\alpha_k = \left\{\begin{array}{ll}
k-1 & \text{if } 1\leq k\leq K \\
KL + t - 1 & \text{if } k = K+t,\ 1\leq t\leq T
\end{array}\right.,\quad
\beta_\ell = \left\{\begin{array}{ll}
K(\ell-1) & \text{if } 1 \leq \ell \leq L \\
KL + t - 1 & \text{if } \ell = L + t,\ 1\leq t \leq T
\end{array}\right.
\end{equation}
if $L \leq K$ and 
\begin{equation}\label{alphabetadefn2 b}
\alpha_\ell = \left\{\begin{array}{ll}
K(\ell-1) & \text{if } 1 \leq \ell \leq L \\
KL + t - 1 & \text{if } \ell = L + t,\ 1\leq t \leq T
\end{array}\right.,\quad
\beta_k = \left\{\begin{array}{ll}
k-1 & \text{if } 1\leq k\leq K \\
KL + t - 1 & \text{if } k = K+t,\ 1\leq t\leq T
\end{array}\right.
\end{equation}
if $K < L$.

Lastly, define $N = |\terms(\alpha\oplus\beta)|$.  Then $\mathsf{GASP}_{\text{big}}$ is defined to be the polynomial code $\mathsf{PC}(K,L,T,N,\alpha,\beta)$.
\end{definition}

\subsection{Decodability and $T$-security}
\begin{theorem} \label{teo: gaspbig secure decodable}
The polynomial code $\mathsf{GASP}_{\text{big}}$ is decodable and $T$-secure.
\end{theorem}
\begin{IEEEproof}
We show that $\alpha\oplus\beta$ is decodable and $T$-secure, and the result then follows from Theorem \ref{blackbox}. If $L \leq K$, then $\alpha$ and $\beta$ are as in (\ref{alphabetadefn2}). Suppose that $k\in [K]$ and $\ell\in [L]$, so that $\alpha_k + \beta_\ell = k-1 + K(\ell-1)$.  As $k$ and $\ell$ range over all of $[K]$ and $[L]$, respectively, each such number gives a unique integer in the interval $[0,KL-1]$.  As every other term in the outer sum $\alpha\oplus\beta$ is greater than or equal to $KL$, we see that the decodability condition of Definition \ref{outersumsecure} is satisfied.  As for $T$-security, it is clear that all $\alpha_{K+t}$ for $t\in [T]$ are distinct, and all $\beta_{L+t}$ for $t\in[T]$ are distinct.  Therefore the $T$-security condition of Definition \ref{outersumsecure} is satisfied. If $K<L$ then the same argument holds by interchanging $\alpha$ and $\beta$.
\end{IEEEproof}

\begin{table}[H]
\centering
 \resizebox{.95\textwidth}{!}{\begin{tabular}{c|ccc|cccc}

 & $\beta_1 = 0$ & $\cdots$ & $\beta_L = K(L-1)$ & \cellcolor{blue!25} $\beta_{L+1} = KL$ & \cellcolor{blue!25} $\beta_{L+2} = KL + 1$ & \cellcolor{blue!25} $\cdots$ & \cellcolor{blue!25}  $\beta_{L+T} = KL + T -1$ \\ 
 \toprule
$\alpha_1 = 0$ & \cellcolor{red!25} $0$ & \cellcolor{red!25} $\cdots$ & \cellcolor{red!25} $K(L-1)$ & $KL$ & $KL+1$ & $\cdots$ & $KL+T-1$ \\
$\vdots$ & \cellcolor{red!25} $\vdots$ & \cellcolor{red!25} $\ddots$ & \cellcolor{red!25} $\vdots$ & $\vdots$ & $\vdots$ & $\ddots$ & $\vdots$ \\
$\alpha_K = K-1$ & \cellcolor{red!25} $K-1$ & \cellcolor{red!25} $\cdots$ & \cellcolor{red!25} $KL-1$ & $KL+K-1$ & $KL+K$ & $\cdots$ & $KL + K + T - 2$\\
\midrule
\cellcolor{green!25} $\alpha_{K+1} = KL$ & $KL$ & $\cdots$ & $2KL-K$ & $2KL$ & $2KL + 1$ & $\cdots$ & $2KL + T - 1$\\
\cellcolor{green!25} $\alpha_{K+2} = KL + 1$ & $KL + 1$ & $\cdots$ & $2KL-K+1$ & $2KL + 1$ & $2KL + 2$ & $\cdots$ & $2KL + T$ \\
\cellcolor{green!25} $\vdots$ & $\vdots$ & $\ddots$ & $\vdots$ & $\vdots$ & $\vdots$ & $\ddots$ & $\vdots$\\
\cellcolor{green!25} $\alpha_{K+T} = KL + T - 1$ & $KL+T-1$ & $\cdots$ & $2KL - K + T - 1$ & $2KL + T - 1$ & $2KL + T$ & $\cdots$ & $2KL + 2T - 2$\\
\bottomrule
\end{tabular}
}\caption{The degree table, $\alpha\oplus\beta$, for the vectors $\alpha$ and $\beta$ as per \eqref{alphabetadefn2}.}\label{bigmatrix2}
\end{table}

\subsection{Download Rate} \label{sec: table division}

To compute the number of terms in the degree table of $\mathsf{GASP}_{\text{big}}$, we divide the table into four regions.

\begin{itemize}
    \item Upper Left: $\mathrm{UL} = \{ (\alpha \oplus \beta)_{ij} : 1\leq i \leq K, 1\leq j \leq  L \}.$
    \item Upper Right: $\mathrm{UR} = \{ (\alpha \oplus \beta)_{ij} : 1\leq i \leq K, L+1\leq j \leq  L+T \}.$
    \item Lower Left: $\mathrm{LL} = \{ (\alpha \oplus \beta)_{ij} : K+1\leq i \leq K+T, 1\leq j \leq  L \}.$
    \item Lower Right: $\mathrm{LR} = \{ (\alpha \oplus \beta)_{ij} : K+1\leq i \leq K+T, L+1\leq j \leq  L+T \}.$
\end{itemize}

Then, we compute the number of terms in each of these regions and use the inclusion-exclusion principle to obtain the number of terms in the whole table.

\begin{theorem}\label{largeNrate}
Let $N = |\terms(\alpha\oplus\beta)|$, where $\alpha$ and $\beta$ are as in Definition \ref{def:gaspbig}. Then $N$ is given by
\begin{equation}\label{largeN}
    N = \left\{\begin{array}{ll}
    (K+T)(L+1) - 1 & \text{if } T< K \\
    2KL + 2T - 1 & \text{if } T\geq K
    \end{array}\right.
\end{equation}
if $L\leq K$, and
\begin{equation}\label{largeN b}
    N = \left\{\begin{array}{ll}
    (L+T)(K+1) - 1 & \text{if } T< L \\
    2KL + 2T - 1 & \text{if } T\geq L
    \end{array}\right.
\end{equation}
if $K<L$.

Consequently, the polynomial code $\mathsf{GASP}_{\text{big}}(K,L,T)$ has rate $\mathcal{R} = KL/N$, where $N$ is as in (\ref{largeN}) or (\ref{largeN b}).
\end{theorem}
\begin{IEEEproof}
The degree table, $\alpha\oplus\beta$, is shown in Table \ref{bigmatrix2}. We first prove for the case where $L \leq K$. We denote by $[A:B]$ the set of all integers in the interval $[A,B]$.  We can describe the terms of the four blocks of $\alpha\oplus\beta$ as follows:
\begin{equation}
    \begin{aligned}
    \terms(\mathrm{UL}) &= [0:KL-1] \\
    \terms(\mathrm{UR}) &= [KL:KL+K+T-2] \\
    \terms(\mathrm{LL}) &= \bigcup_{\ell = 0}^{L-1} [KL+K(\ell-1):KL+K(\ell-1)+T-1] \\
    \terms(\mathrm{LR}) &= [2KL:2KL+2T-2]
    \end{aligned}
\end{equation}
The sizes of these sets is given by
\begin{equation}
    \begin{aligned}
    |\terms(\mathrm{UL})| &= KL \\
    |\terms(\mathrm{UR})| &= K + T - 1 \\
    |\terms(\mathrm{LL})| &= \left\{\begin{array}{ll}
    LT & \text{if } T\leq K \\
    KL - K + T & \text{if } T\geq K
    \end{array}\right.\\
    |\terms(\mathrm{LR})| &= 2T-1
    \end{aligned}
\end{equation}
Since the largest term in $\mathrm{UL}$ is smaller than any term on the other blocks, $\terms(\mathrm{UL})$ is disjoint from the terms of the other blocks.  One then observes that the pairwise intersections of the sets of terms of the blocks are given by
\begin{equation}
    \begin{aligned}
    \terms(\mathrm{UR})\cap\terms(\mathrm{LL}) &= 
    \left\{\begin{array}{ll}
       [K:K+T-1]  & \text{if } L = 1 \\
       
       [KL:KL+T-1]\cup [KL+K:KL+K+T-2] & \text{if } L\geq 2
    \end{array}\right. \\
    \terms(\mathrm{LL})\cap\terms(\mathrm{LR}) &= [2KL:2KL-K+T-1] \\
    \terms(\mathrm{UR})\cap\terms(\mathrm{LR}) &= [2KL:KL+K+T-2]
    \end{aligned}
\end{equation}
The sizes of these pairwise intersections are now calculated to be
\[
\begin{aligned}
|\terms(\mathrm{UR})\cap\terms(\mathrm{LL})| &= \left\{\begin{array}{ll}
T & \text{if } L = 1 \\
2T - 1 & \text{if } L\geq 2, T\leq K \\
K+T-1 & \text{if } L\geq 2, T\geq K
\end{array}\right. \\
|\terms(\mathrm{LL})\cap\terms(\mathrm{LR})| &= \left\{\begin{array}{ll}
0 & T\leq K \\
T-K & T\geq K
\end{array}\right. \\
|\terms(\mathrm{UR})\cap\terms(\mathrm{LR})| &=
\left\{\begin{array}{ll}
0 & T\leq K(L-1)+1 \\
T-(K(L-1)+1) & T\geq K(L-1)+1
\end{array}\right.
\end{aligned}
\]
Finally, the triple intersection is given by
\[
\terms(\mathrm{UR})\cap\terms(\mathrm{LL})\cap\terms(\mathrm{LR}) = [2KL:\min\{2KL-K+T-1,KL+K+T-2\}]
\]
We have $2KL-K+T-1\leq KL+K+T-2$ if and only if $L = 1$.  One now computes that
\[
|\terms(\mathrm{UR})\cap\terms(\mathrm{LL})\cap\terms(\mathrm{LR})| = \left\{\begin{array}{ll}
0 & \text{if } L = 1, T\leq K \\
T-K & \text{if } L = 1, T\geq K \\
0 & \text{if } L\geq 2, T\leq K(L-1)+1 \\
T-(K(L-1)+1) & \text{if } L\geq 2, T\geq K(L-1)+1
\end{array}\right.
\]
We can now compute $N = |\terms(\alpha\oplus\beta)|$ by using the inclusion-exclusion principle, as
\[
\begin{aligned}
N = |\terms(\alpha\oplus\beta)| &= |\terms(\mathrm{UL})| + |\terms(\mathrm{UR})| + |\terms(\mathrm{LL})| + |\terms(\mathrm{LR})| \\
&- |\terms(\mathrm{UR})\cap\terms(\mathrm{LL})| - |\terms(\mathrm{LL})\cap\terms(\mathrm{LR})| - |\terms(\mathrm{UR})\cap\terms(\mathrm{LR})| \\
&+ |\terms(\mathrm{UR})\cap\terms(\mathrm{LR})\cap\terms(\mathrm{LL})|.
\end{aligned}
\]

For $K < L$, the proof is analogous by interchanging $\alpha$ and $\beta$.
\end{IEEEproof}

\begin{remark}
If we take $K = L = 1$, then the polynomial code $\mathsf{GASP}_{\text{big}}(1,1,T)$ uses $N = 2T+1$ servers.  Thus for any $N$ we can construct a polynomial code for any $T\leq \left\lfloor\frac{N-1}{2}\right\rfloor$, which is the same range of allowable $T$ as in \cite{kakar}.
\end{remark}

\subsection{Performance}

We now compare $\mathsf{GASP}_{\text{big}}$ with the polynomial codes of \cite{ravi2018mmult} and \cite{kakar}. Indeed, we show that it outperforms them for most parameters. To do this it suffices, because of \eqref{eq:kakar rate}, to show that $N$, as defined in Theorem \ref{largeNrate}, is smaller than $(K+T)(L+1)-1$.

\begin{theorem}
Let $N$ be defined as in Theorem \ref{largeNrate}. Then, $N \leq (K+T)(L+1)-1$.
\end{theorem}

\begin{IEEEproof}
Suppose $L \leq K$. Then $N$ is as in \eqref{largeN}. We will analyze each case.

\begin{itemize}
    \item If $T<K$: then $(K+T)(L+1)-1 = N$.
    \item If $T\geq K$: then $0 \leq (L-1)(T-K) = ((K+T)(L+1)-1) - (2KL + 2T -1)$. 
    
    Thus, $(K+T)(L+1)-1 \geq 2KL + 2T -1 = N$.
\end{itemize}
The result for $K < L$ follows by switching the roles of $K$ and $L$ in the above calculation.

    
    

\end{IEEEproof}

\begin{figure}[H]
    \centering
    \includegraphics[width=.4\textwidth]{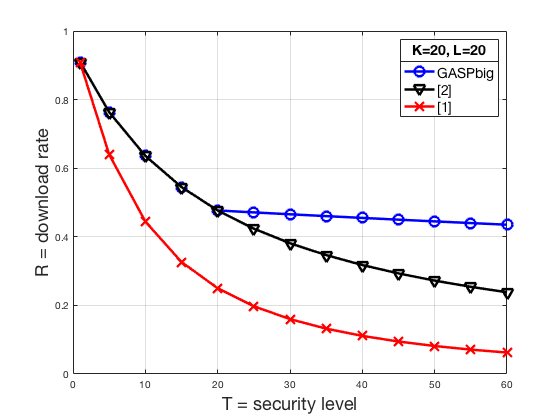} \hspace{2.5em}
    \includegraphics[width=.4\textwidth]{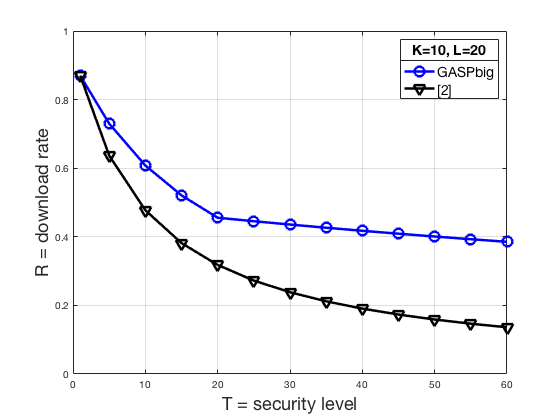} \\
    \caption{Comparison of the Polynomial Code $\mathsf{GASP}_{\text{big}}$ with that of \cite{kakar}.  We plot the rate of both schemes for $K = L=20$ on the left, and $K = 10, L = 20$ on the right.}
    \label{fig:compare2}
\end{figure}

\section{A Polynomial Code for Small $T$} \label{sec:gaspsmall}

In this section, we construct a polynomial code $\mathsf{GASP}_{\text{small}}$ which outperforms $\mathsf{GASP}_{\text{big}}$ when $T< \min \{K,L\}$.  This is done by choosing the $\alpha_k$ and $\beta_\ell$ to lie in certain arithmetic progressions so that the columns of the upper-right block of $\alpha\oplus\beta$, shown in Table \ref{bigmatrix}, overlap as much as possible, and similarly for the columns of the lower-left block.  For $T$ small relative to $K$ and $L$, these two blocks are much bigger than the lower-right block, which the scheme construction essentially ignores.

\begin{table}[!htb]
\centering
 \resizebox{.95\textwidth}{!}{\begin{tabular}{c|ccc|cccc}

 & $\beta_1 = 0$ & $\cdots$ & $\beta_L = K(L-1)$ & \cellcolor{blue!25} $\beta_{L+1} = KL$ & \cellcolor{blue!25} $\beta_{L+2} = KL + 1$ & \cellcolor{blue!25} $\cdots$ & \cellcolor{blue!25} $\beta_{L+T} = KL + T -1$ \\ 
\toprule
$\alpha_1 = 0$ & \cellcolor{red!25} $0$ & \cellcolor{red!25} $\cdots$ & \cellcolor{red!25} $K(L-1)$ & $KL$ & $KL+1$ & $\cdots$ & $KL+T-1$ \\
$\vdots$ & \cellcolor{red!25} $\vdots$ & \cellcolor{red!25} $\ddots$ & \cellcolor{red!25} $\vdots$ & $\vdots$ & $\vdots$ & $\ddots$ & $\vdots$ \\
$\alpha_K = K-1$ & \cellcolor{red!25} $K-1$ & \cellcolor{red!25} $\cdots$ & \cellcolor{red!25} $KL-1$ & $KL+K-1$ & $KL+K$ & $\cdots$ & $KL + K + T - 2$\\
\midrule
\cellcolor{green!25} $\alpha_{K+1} = KL$ & $KL$ & $\cdots$ & $2KL-K$ & $2KL$ & $2KL + 1$ & $\cdots$ & $2KL + T - 1$\\
\cellcolor{green!25} $\alpha_{K+2} = KL + K$ & $KL +K$ & $\cdots$ & $2KL$ & $2KL + K$ & $2KL + K + 1$ & $\cdots$ & $2KL + K + T - 1$ \\
\cellcolor{green!25} $\vdots$ & $\vdots$ & $\ddots$ & $\vdots$ & $\vdots$ & $\vdots$ & $\ddots$ & $\vdots$\\
\cellcolor{green!25} $\alpha_{K+T} = KL + K(T-1)$ & $KL+K(T-1)$ & $\cdots$ & $2KL + K(T-2)$ & $2KL + K(T-1)$ & $2KL + K(T-1)+1$ & $\cdots$ & $2KL + (K+1)(T-1)$\\
\bottomrule
\end{tabular}
}\caption{The degree table, $\alpha\oplus\beta$, of the vectors $\alpha$ and $\beta$ as per Definition \ref{def:gaspsmall}.}\label{bigmatrix}
\end{table}

\begin{definition} \label{def:gaspsmall}
Given $K$, $L$, and $T$, define the polynomial code  $\mathsf{GASP}_{\text{small}}$ as follows. Let $\alpha$ and $\beta$ be given by
\begin{equation}\label{alphabetadefn}
\alpha_k = \left\{\begin{array}{ll}
k-1 & \text{if } 1\leq k\leq K \\
KL + K(t-1) & \text{if } k = K+t,\ 1\leq t\leq T
\end{array}\right.,\quad
\beta_\ell = \left\{\begin{array}{ll}
K(\ell-1) & \text{if } 1\leq \ell\leq L \\
KL + t-1 & \text{if } \ell = L+t,\ 1\leq t\leq T.
\end{array}\right.
\end{equation}
if $K\leq L$, and
\begin{equation}\label{alphabetadefn}
\alpha_\ell= \left\{\begin{array}{ll}
K(\ell-1) & \text{if } 1\leq \ell\leq L \\
KL + t-1 & \text{if } \ell = L+t,\ 1\leq t\leq T.
\end{array}\right.,\quad
\beta_k = \left\{\begin{array}{ll}
k-1 & \text{if } 1\leq k\leq K \\
KL + K(t-1) & \text{if } k = K+t,\ 1\leq t\leq T
\end{array}\right.
\end{equation}
if $L<K$.

Lastly, define $N = |\terms(\alpha\oplus\beta)|$.  Then $\mathsf{GASP}_{\text{small}}$ is defined to be the polynomial code $\mathsf{PC}(K,L,T,N,\alpha,\beta)$.
\end{definition}
The example in Section \ref{sec:MainExample} is exactly the polynomial code $\mathsf{GASP}_{\text{small}}$ when $K=L=3$ and $T=2$.  In what follows we show that $\mathsf{GASP}_{\text{small}}$ is decodable and $T$-secure, and compute its download rate.  Throughout this section, $\alpha$ and $\beta$ will be as in Definition \ref{def:gaspsmall}.

\subsection{Decodability and $T$-security}

\begin{theorem}
The polynomial code $\mathsf{GASP}_{\text{small}}(K,L,T)$ is decodable and $T$-secure.  
\end{theorem}
\begin{IEEEproof}
Analogous to Theorem \ref{teo: gaspbig secure decodable}.
\end{IEEEproof}

\subsection{Download Rate} 

We now find the download rate of $\mathsf{GASP}_{\text{small}}$ by computing $N=|\terms(\alpha\oplus\beta)|$.

\begin{theorem}\label{count}
Let $N = |\terms(\alpha\oplus\beta)|$, where $\alpha$ and $\beta$ are as in Definition \ref{def:gaspsmall}. Then $N$ is given by
\begin{equation}\label{smallN}
N = \left\{\begin{array}{ll}
2K + T^2 & \text{if } L = 1, T< K \\
KT+K+T & \text{if } L = 1, T\geq K \\
KL + K + L & \text{if } L\geq 2, 1 = T< K \\
KL + K + L + T^2 + T - 3 & \text{if } L\geq 2, 2\leq T < K \\
KL + KT + L + 2T - 3 - \left\lfloor\frac{T-2}{K}\right\rfloor & \text{if } L \geq 2, K\leq T\leq K(L-1)+1 \\
2KL + KT-K+T & \text{if } L\geq 2, K(L-1)+1 \leq T
\end{array}\right.
\end{equation}
if $K \leq L$, and
\begin{equation}\label{smallN b}
N = \left\{\begin{array}{ll}
2L + T^2 & \text{if } K = 1, T< L \\
LT+L+T & \text{if } K = 1, T\geq L \\
KL + K + L & \text{if } K\geq 2, 1 = T< L \\
KL + K + L + T^2 + T - 3 & \text{if } K\geq 2, 2\leq T < L \\
KL + LT + K + 2T - 3 - \left\lfloor\frac{T-2}{L}\right\rfloor & \text{if } K \geq 2, L\leq T\leq L(K-1)+1 \\
2KL + LT-L+T & \text{if } K\geq 2, L(K-1)+1 \leq T
\end{array}\right.
\end{equation}
if $L<K$.

Consequently, the polynomial code $\mathsf{GASP}_{\text{small}}$ has rate $\mathcal{R} = KL/N$, where $N$ is as in (\ref{smallN}) or (\ref{smallN b}).
\end{theorem}

\begin{IEEEproof}
The proof is in Appendix \ref{apen: theorem 6}.
\end{IEEEproof}

\begin{remark}
 The above construction for small $T$ is from where the name GASP (Gap Additive Secure Polynomial) is derived.  The construction allows for gaps in the degrees of monomials appearing as summands of $h(x)$, as was observed in the example in Section \ref{sec:MainExample}.  Allowing for these gaps gives one more flexibility in how the vectors $\alpha$ and $\beta$ are chosen to attempt to minimize $N = |\terms(\alpha\oplus\beta)|$.  Note that for very large $T$, the inequality $T\geq K(L-1) + 1$ has forced the outer sum $\alpha\oplus\beta$ to contain every integer from $0$ to $2KL + (K+1)(T-1)$, with no gaps.
\end{remark}

\subsection{Performance}

We now show that $\mathsf{GASP}_{\text{small}}$ outperforms $\mathsf{GASP}_{\text{big}}$ when $T < \min \{K,L \}$. 

\begin{theorem} \label{teo:smallbigcompare}
Let $T < \min \{K,L \}$. Then $N_\text{small} \leq N_\text{big}$.
\end{theorem}

\begin{IEEEproof}
We will analyze each case.

\begin{itemize}
    \item If $T=1$: then $N_\text{big} = KL + K + L = N_\text{small}$. 
    \item If $T\geq 2$: then $0 \leq (T+1)(T-1) - T^2 + 2 \leq L(T-1) - T^2 + 2 = N_\text{big} -  N_\text{small}$.
    
    Thus, $N_\text{big} \geq N_\text{small}$.

\end{itemize}

\end{IEEEproof}

\begin{figure}[H]
    \centering
    \includegraphics[width=.39\textwidth]{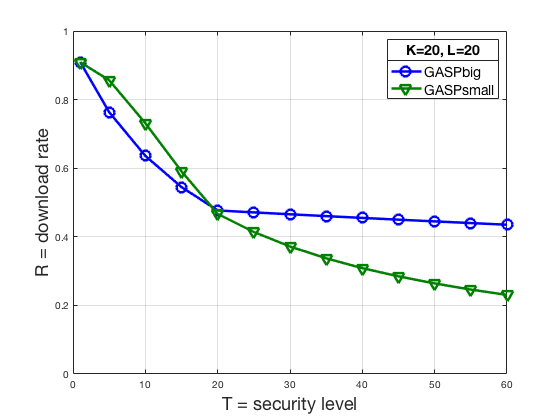} \hspace{2.5em}
    \includegraphics[width=.39\textwidth]{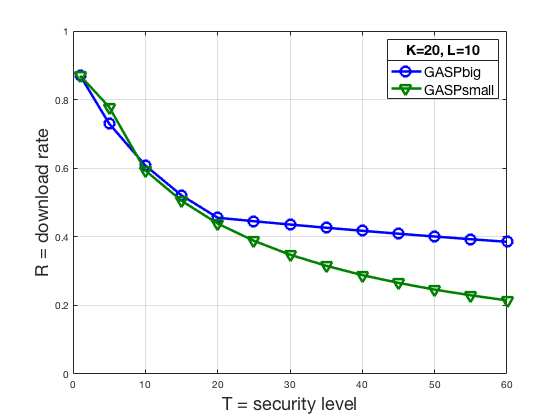} \\
    \caption{Comparison between $\mathsf{GASP}_{\text{small}}$ and $\mathsf{GASP}_{\text{big}}$.  We plot the rate of both schemes for $K = 20$ and $L = 20$ on the left, and $K = 20$ and $L = 10$ on the right. As shown, $\mathsf{GASP}_{\text{small}}$ outperforms $\mathsf{GASP}_{\text{big}}$ for $T < \max \{K,L \}$.}
    \label{fig:compare3}
\end{figure}

\section{Combining Both Schemes}\label{sec:combining}

In this section, we construct a polynomial, $\mathsf{GASP}$, by combining both $\mathsf{GASP}_{\text{small}}$ and $\mathsf{GASP}_{\text{big}}$. By construction, $\mathsf{GASP}$ has a better rate than all previous schemes.

\begin{definition}
Given $K$, $L$, and $T$, we define the polynomial code $\mathsf{GASP}$ to be
\begin{equation}
    \mathsf{GASP} = \left\{\begin{array}{ll}
    \mathsf{GASP}_{\text{small}} & \text{if } T < \min \{K,L \} \\
    \mathsf{GASP}_{\text{big}} & \text{if } T \geq \min \{K,L \}.
    \end{array}\right.
\end{equation}
\end{definition}

\begin{theorem}
For $L \leq K$, the polynomial code $\mathsf{GASP}$ has rate,
\[
\mathcal{R} = \left\{\begin{array}{cl}
\dfrac{KL}{KL+K+L} & \text{if } 1=T<L\leq K \\[13pt]
\dfrac{KL}{KL+K+L+T^2+T-3} & \text{if } 2\leq T < L \leq K \\[13pt]
\dfrac{KL}{(K+T)(L+1) - 1} & \text{if } L \leq T < K\\[13pt]
\dfrac{KL}{2KL + 2T - 1} & \text{if } L \leq K \leq T
\end{array}\right. 
\]
For $K<L$, the rate is given by interchanging $K$ and $L$.
\end{theorem}

\begin{IEEEproof}
Follows immediately from Theorems \ref{largeNrate}, \ref{count}, and \ref{teo:smallbigcompare}.
\end{IEEEproof}

\subsection{Fixed Computation Load}

We now compare the rate of $\mathsf{GASP}$ with those of \cite{ravi2018mmult} and \cite{kakar} when $K$ and $L$ are fixed.  Throughout this section, we let
\begin{equation}
    \mathcal{R}_1 = \frac{K^2}{(K+T)^2}\quad\text{and}\quad \mathcal{R}_2 = \frac{KL}{(K+T)(L+1)-1}.
\end{equation}
Here $\mathcal{R}_1$ and $\mathcal{R}_2$ are the rates of the polynomial codes in \cite{ravi2018mmult} and \cite{kakar}, respectively.


\begin{figure}[H]
    \centering
    \includegraphics[width=.4\textwidth]{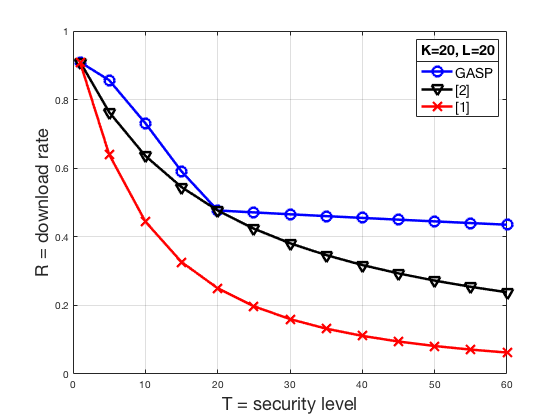} \hspace{2.5em}
    \includegraphics[width=.4\textwidth]{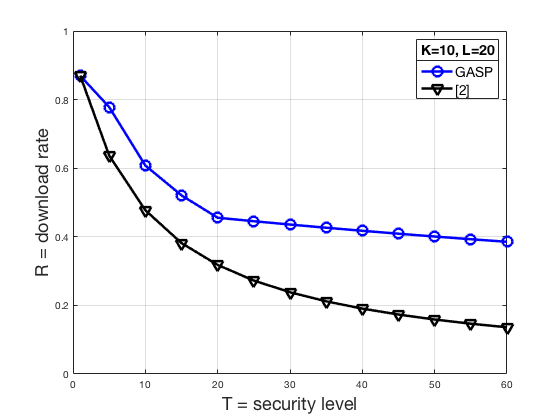} \\
    \caption{Comparison of the Polynomial Code $\mathsf{GASP}$ with that of \cite{ravi2018mmult} and \cite{kakar}. We plot the rate of the schemes for $K = 20$ and $L = 20$ on the left, and $K = 10$ and $L = 20$ on the right.}
    \label{fig:compare2}
\end{figure}



\subsection{Fixed number of workers}\label{workers}
To deepen the comparison with \cite{kakar}, we plot the download rates $\mathcal{R}$ and $\mathcal{R}_2$ as functions of the total number $N$ of servers and the security level $T$.  For $\mathsf{GASP}$ and the polynomial code of \cite{kakar}, given some $N$ and $T$, we must calculate a $K$ and $L$ for which the expression for the required number of servers is less than the given $N$, and which ideally maximizes the rate function.  In \cite[Theorem 1]{kakar}, the authors propose the solution
\begin{equation}\label{kakarkl}
\hat{L} = \max\left\{1,\left\lceil -\frac{3}{2} +\sqrt{\frac{1}{4} + \frac{N}{T}}\right\rceil\right\},\quad \hat{K} = \left\lfloor \frac{N+1}{\hat{L}+1}-T\right\rfloor
\end{equation}
which, for a given $N$ and $T$, is shown to satisfy $(\hat{K}+T)(\hat{L}+1)-1\leq N$ and nearly maximize the rate function $\mathcal{R}_1$.

For a given $N$ and $T$, optimizing the rate of $\mathsf{GASP}$ presents one with the following optimization problem:
\begin{equation}\label{toohard}
\begin{aligned}
\max_{K,L}\quad & \mathcal{R}_{\max} = \frac{KL}{\min\{N_{\text{small}},N_{\text{big}}\}} \\
\text{subject to}\quad & \min\{N_{\text{small}},N_{\text{big}}\}\leq N
\end{aligned}
\end{equation}
Due to the complicated nature of the expressions for $N_{\text{small}}$ and $N_{\text{big}}$, we will not attempt to solve this optimization problem analytically.  Instead, for the purposes of the present comparison with \cite{kakar}, we simply solve (\ref{toohard}) by brute force for each specific value of $N$ and $T$.

In Fig.\ \ref{fig:compare3} we plot the download rate of $\mathsf{GASP}$ versus the download rate of the polynomial code of \cite{kakar}, for $N = 50$ and $N = 100$ servers.  The optimal values of $K$ and $L$ for $\mathsf{GASP}$ were computed by solving (\ref{toohard}) by brute force.  The values of $K$ and $L$ for the scheme of \cite{kakar} were those of (\ref{kakarkl}).

\begin{figure}[H]
    \centering
    \includegraphics[width=.4\textwidth]{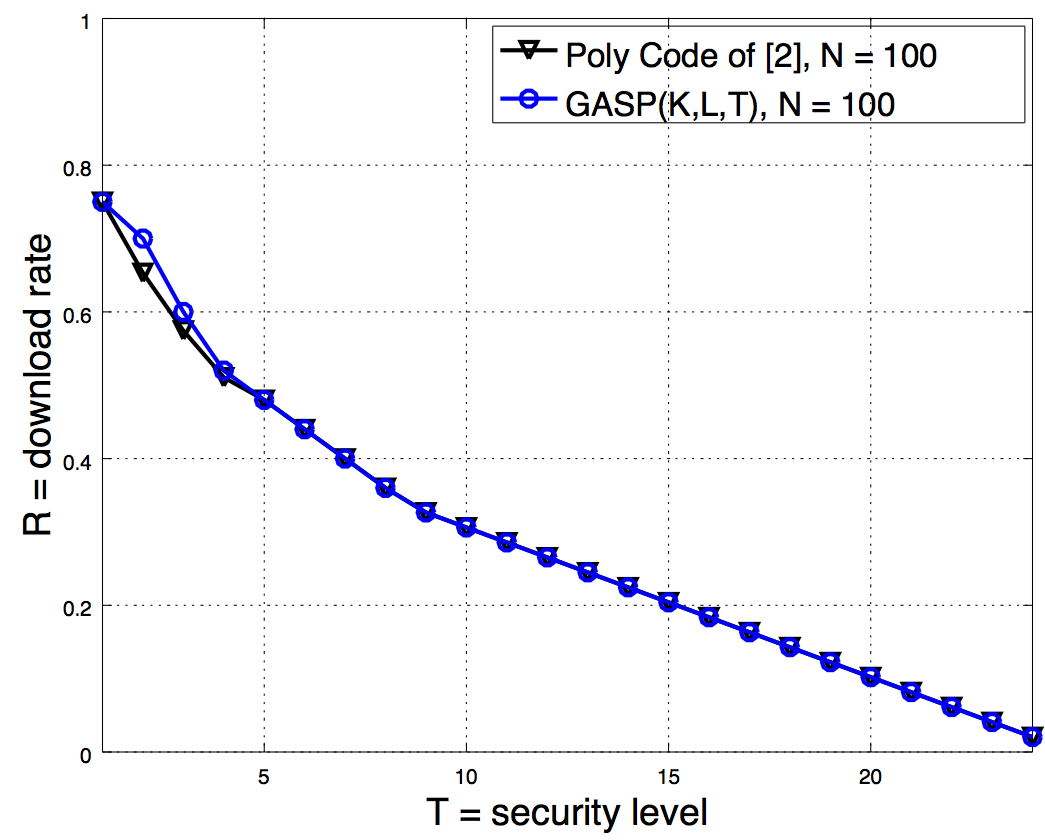}
    \caption{The rate of $\mathsf{GASP}$ and of the the polynomial code of \cite{kakar}, as a function of the level of security $T$, for $N = 50$ servers. The optimal rate of $\mathsf{GASP}$ for a given $N$ and $T$ was computed by finding a solution to (\ref{toohard}) by brute force.}
    \label{fig:compare3}
\end{figure}

\update{The apparent equality in rate of the two schemes outside of the `small $T$' regime can be explained as follows.  One can show easily that when $T>N/6$ we have $\hat{L} = 1$, and hence the rate from \cite{kakar} is given by $\mathcal{R}_2 = \frac{\hat{K}}{2\hat{K}+2T-1}$, where $\hat{K} = \left\lfloor \frac{N+1}{2}-T\right\rfloor$.  Now the rate of $\mathsf{GASP}$ in this regime is that of $\mathsf{GASP}_{\text{big}}$, so $\mathcal{R} = \frac{KL}{2KL + 2T - 1}$.  Optimizing the rate of $\mathsf{GASP}_{\text{big}}$ for fixed $N$ is now simply a matter of picking the optimal value of $KL$.  Whatever this optimal value happens to be, it only depends on the product $KL$ and not the individual values of $K$ and $L$.  So when optimizing the rate of $\mathsf{GASP}_{\text{big}}$ for fixed $N$ and $T$, one is free to set $L = 1$ without loss of generality.  The rates of the $\mathsf{GASP}_{\text{big}}$ and the scheme of \cite{kakar} are then easily seen to agree.}

\section{Acknowledgements}

\update{We kindly thank the authors of \cite{kakar} for pointing out errors in the plots of Fig.\ \ref{fig:compare3} in the original version of this paper.  These plots have since been corrected.}




\appendix

\subsection{Proof of Theorem \ref{blackbox}}

We will require the following definition throughout the proof of Theorem \ref{blackbox}.

\begin{definition}
Let $\F_q$ be a finite field, let ${\bf a} = (a_1,\ldots,a_N)\in \F_q^N$,  and let $\mathcal{J}$ be a set of non-negative integers of size $|\mathcal{J}| = N$.  We define the \emph{Generalized Vandermonde Matrix} $GV({\bf a},\mathcal{J})\in\F_q^{N\times N}$ to be
\[
GV({\bf a},\mathcal{J}) = \left[a_n^j\right],\quad 1\leq n\leq N,\ j\in\mathcal{J}.
\]
\end{definition}
Note that if $\mathcal{J} = \{0,1,\ldots,N-1\}$ and the $a_n$ are all chosen distinct, then $GV({\bf a},\mathcal{J})$ is the familiar $N\times N$ Vandermonde matrix associated with the $a_n$, and is invertible if and only if the $a_n$ are distinct.

We begin proving Theorem \ref{blackbox} by stating the following useful Lemma.  For all practical purposes, this reduces checking decodability and $T$-security to checking polynomial conditions.  We say a matrix has the \emph{MDS property} if every maximal minor has non-zero determinant.  Equivalently, the matrix is the generator matrix of an MDS code.

\begin{lemma}\label{decodabilityandsecurity}
Let $\mathsf{PC}(K,L,T,\alpha,\beta)$ be a polynomial code, such that $\alpha\oplus\beta$ is decodable and $T$-secure.  Suppose that there is an evaluation vector ${\bf a} = (a_1,\ldots,a_N)\in\F_{q^r}^N$ such that the following properties hold:
\begin{itemize}
    \item[(i)] (Decodability) The Generalized Vandermonde Matrix $GV({\bf a},\mathcal{J})$ is invertible.
    \item[(ii)] ($T$-privacy) The $T\times N$ matrices
\[
P = \begin{bmatrix}
a_n^{\alpha_{K+t}}
\end{bmatrix}\quad\text{and}\quad
Q = \begin{bmatrix}
a_n^{\beta_{L+t}}
\end{bmatrix},
\]
where $1\leq t\leq T$ and $1\leq n\leq N$,
have the MDS property.  
\end{itemize}
Then $\mathsf{PC}(K,L,T,\alpha,\beta)$ is decodable and $T$-secure.
\end{lemma}

\begin{IEEEproof}
Since the matrix $GV({\bf a},\mathcal{J})$ is invertible, the polynomial $h(x) = \sum_{j\in\mathcal{J}}C_jx^j$ can be interpolated from the evaluations $h(a_n)$, for $n = 1,\ldots,N$.  Thus the user can recover all of the coefficients of $h(x)$.  By the decodability condition of the outer sum $\alpha\oplus\beta$, the user can then recover all products $A_kB_\ell$.

\update{The argument for $T$-privacy is familiar and follows the proof of $T$-security in Equation (28) in the proof of Theorem 2 in \cite{ravi2018mmult}.}  One shows that, given the above condition, any $T$-tuple of matrices $f(a_{n_1}),\ldots,f(a_{n_T})$ is uniform random on the space of all $T$-tuples of matrices of the appropriate size, and is independent of $A$.  The same argument works for $B$.
\end{IEEEproof}

Let us now finish the proof of Theorem \ref{blackbox}.  Let ${\bf X} = (X_1,\ldots,X_N)$ be a vector of variables and consider the polynomial
\begin{equation}
    D({\bf X}) = \det(GV({\bf X},\mathcal{J})).
\end{equation}
Additionally, if $\mathcal{T} = \{n_1,\ldots,n_T\}\subseteq[N]$ is any set of size $T$, define
\begin{equation}
    P_{\mathcal{T}}({\bf X}) = \det\begin{bmatrix}
    X_{n_t}^{\alpha_{K+t}}
    \end{bmatrix}
    \quad\text{and}\quad
    Q_{\mathcal{T}}({\bf X}) = 
    \det\begin{bmatrix}
    X_{n_t}^{\beta_{L+t}}
    \end{bmatrix}.
\end{equation}
By Lemma \ref{decodabilityandsecurity}, it suffices to find an evaluation vector ${\bf a}\in\F_{q^r}^N$ such that $D({\bf a})\neq 0$, $P_{\mathcal{T}}({\bf a})\neq 0$, and $Q_{\mathcal{T}}({\bf a})\neq 0$ for all $\mathcal{T}\subseteq[N]$ of size $T$.  By the assumption that $\alpha\oplus\beta$ is decodable and $T$-secure, none of the polynomials $D({\bf X})$, $P_{\mathcal{T}}({\bf X})$, and $Q_{\mathcal{T}}({\bf X})$ are zero, and all have degree bounded by $J:= \sum_{j\in\mathcal{J}}j$.

Now consider a finite extension $\F_{q^r}$ of $\F_q$ and a subset $G\subseteq \F_{q^r}$ of size $G > \left(2\binom{N}{T}+1\right)J$.  Sample each entry $a_n$ of ${\bf a}$ uniformly at random from $G$.  Let $\mathsf{E}$ be the union of the events $D({\bf a}) = 0$, $P_{\mathcal{T}}({\bf a}) = 0$, and $Q_{\mathcal{T}}({\bf a}) = 0$ for all $\mathcal{T}\subseteq[N]$ of size $T$. To finish the proof, it suffices to show that $\text{Pr}(\mathsf{E}) < 1$.  By the union bound and the Schwarz-Zippel Lemma, we have
\[
\text{Pr}(\mathsf{E}) \leq \text{Pr}(D({\bf a}) = 0) + \sum_{\substack{ \mathcal{T}\subseteq[N]\\|\mathcal{T}| = T }} \text{Pr}(P_{\mathcal{T}}({\bf a}) = 0) + \sum_{\substack{ \mathcal{T}\subseteq[N]\\|\mathcal{T}| = T }} \text{Pr}(Q_{\mathcal{T}}({\bf a}) = 0) 
\leq \left(2\binom{N}{T}+1\right)\frac{J}{G} < 1.
\]
This completes the proof of the Theorem. \hfill $\blacksquare$

\subsection{Proof of Theorem \ref{count}.} \label{apen: theorem 6}

The degree table, $\alpha\oplus\beta$, is is shown in Table \ref{bigmatrix}. We first prove for the case where $L \leq K$.  As in section \ref{sec: table division}, we let $\mathrm{UL}$, $\mathrm{UR}$, $\mathrm{LL}$, and $\mathrm{LR}$ be the upper-left, upper-right, lower-left, and lower-right blocks, respectively, of $\alpha\oplus\beta$. We first count the number in each block, and then study the intersections of the blocks.  

It will be convenient to adopt the following notation.  For integers $A$, $B$, and $C$, let $[A:B~|~C]$ be the set of all multiples of $C$ in the interval $[A,B]$.  If $A = DC$ is a multiple of $C$, we have 
\[
|[DC:B~|~C]| = \left(\left\lfloor \frac{B}{C}\right\rfloor-D+1\right)^+\]
where $x^+ = \max\{x,0\}$.  If $C = 1$ then we write $[A:B]$ instead of $[A:B~|~1]$, so that $[A:B]$ denotes all the integers in the interval $[A,B]$.

The sets $\terms(\mathrm{UL})$, $\terms(\mathrm{UR})$, $\terms(\mathrm{LL})$, and $\terms(\mathrm{LR})$ are given by
\begin{equation}
    \begin{aligned}
    \terms(\mathrm{UL}) &= [0:KL-1] \\
    \terms(\mathrm{UR}) &= [KL:KL+K+T-2] \\
    \terms(\mathrm{LL}) &= [KL:2KL+K(T-2)~|~K] \\
    \terms(\mathrm{LR}) &= \bigcup_{t = 1}^{T}\ [2KL+K(t-1):2KL+K(t-1)+T-1] \\
    \end{aligned}
\end{equation}
From these expressions, one can count the sizes of the above sets to be
\begin{equation}
    \begin{aligned}
        |\terms(\mathrm{UL})| &= KL \\
        |\terms(\mathrm{UR})| &= K+T-1 \\
        |\terms(\mathrm{LL})| &= L+T-1 \\
        |\terms(\mathrm{LR})| &= \left\{\begin{array}{ll}
        T^2 & \text{if } T < K \\
        KT-K+T & \text{if } T\geq K
        \end{array}\right.\\
    \end{aligned}
\end{equation}
To understand the last expression above, note that the intervals in the union expression for $\terms(\mathrm{LR})$ consist of all integers from $2KL$ to $2KL+(K+1)(T-1)$ exactly when $T\geq K$.

As for intersections, clearly $\terms(\mathrm{UL})$ intersects none of the sets of terms from the other blocks.  Thus it suffices to understand the pairwise intersections among the other three blocks, and the triple intersection of the other three blocks.  Two of these pairwise intersections and their sizes are easily understood:
\begin{equation}
    \begin{aligned}
    \terms(\mathrm{UR})\cap\terms(\mathrm{LL}) &= [KL:KL+K+T-2~|~K], & |\terms(\mathrm{UR})\cap\terms(\mathrm{LL})| &= \left\lfloor \frac{T-2}{K}\right\rfloor + 2\\
    \terms(\mathrm{LL})\cap\terms(\mathrm{LR}) &= [2KL:2KL+K(T-2)~|~K], & |\terms(\mathrm{LL})\cap\terms(\mathrm{LR})| &= T-1
    \end{aligned}
\end{equation}
Understanding the intersection $\terms(\mathrm{UR})\cap\terms(\mathrm{LR})$ is a bit more subtle, and we break the problem into two cases.  If $T\geq K$, then $\terms(\mathrm{LR})=[2KL:2KL+(K+1)(T-1)]$.  Since $\terms(\mathrm{UR})=[KL:KL+K+T-2]$, we see that $\terms(\mathrm{UR})\cap\terms(\mathrm{LR})=[2KL:KL+K+T-2]$.  In the case $T<K$ we have $2KL>KL+K+T-2$ and thus the intersection is empty, unless $L = 1$, in which case $\terms(\mathrm{UR})\cap\terms(\mathrm{LR})=[2K:2K+T-2]$.  It follows that
\begin{equation}\label{doubleintersection}
|\terms(\mathrm{UR})\cap\terms(\mathrm{LR})|
= \left\{\begin{array}{ll}
T-1 & \text{if } T < K, L = 1 \\
0 & \text{if } T < K, L \geq 2 \\
(T-(K(L-1)+1))^+ & \text{if } T\geq K
\end{array}\right.
\end{equation}

It remains to count the size of the triple intersection.  First suppose that $T<K$, which we break into two subcases: (i) $L = 1$ and (ii) $L \geq 2$.  If $L =1$ and $T = 1$, then the triple intersection is empty, but if $T> 1$ then all three blocks intersect in the lone terms $2K$.  If $L\geq 2$, then the triple intersection is again empty by the above paragraph.  Now suppose that $T\geq K$.  In this case the intersection is the set  $[2KL:KL+K+T-2~|~K]$, which has size $\left(
\left\lfloor\frac{T-2}{K}\right\rfloor-L+2
\right)^+$.  We therefore have
\begin{equation}\label{tripleintersection}
|\terms(\mathrm{UR})\cap\terms(\mathrm{LR})\cap\terms(\mathrm{LL})|
= \left\{\begin{array}{ll}
0 & \text{if } L = 1, 1 = T < K\\
1 & \text{if } L = 1, 2\leq T<K \\
0 & \text{if } L\geq2, T<K \\
\left(
\left\lfloor\frac{T-2}{K}\right\rfloor -L+2
\right)^+ & \text{if } T\geq K
\end{array}
\right.
\end{equation}

We can now compute $N = |\terms(\alpha\oplus\beta)|$ by using the inclusion-exclusion principle, as
\[
\begin{aligned}
N = |\terms(\alpha\oplus\beta)| &= |\terms(\mathrm{UL})| + |\terms(\mathrm{UR})| + |\terms(\mathrm{LL})| + |\terms(\mathrm{LR})| \\
&- |\terms(\mathrm{UR})\cap\terms(\mathrm{LL})| - |\terms(\mathrm{LL})\cap\terms(\mathrm{LR})| - |\terms(\mathrm{UR})\cap\terms(\mathrm{LR})| \\
&+ |\terms(\mathrm{UR})\cap\terms(\mathrm{LR})\cap\terms(\mathrm{LL})|.
\end{aligned}
\]
The above computation is straightforward given that we have already calculated the sizes of each of the individual sets.  The only subtlety in deriving the formula (\ref{smallN}) for $N$ arises in the case that $L\geq 2$ and $K(L-1)+1\leq T$.  In this case, one uses the fact that
\[
T-K(L-1)+1\geq 0\Leftrightarrow \left\lfloor \frac{T-2}{K}\right\rfloor -L + 2 \geq 0.
\]
From this equivalence and equations (\ref{doubleintersection}) and (\ref{tripleintersection}) one can use inclusion-exclusion to compute the value of $N$.

For $K < L$, the proof is analogous by interchanging $\alpha$ and $\beta$.

This completes the proof of the Theorem.
\black

\subsection{A Note on the Communication Rate}

The results in this paper were presented in terms of the download rate, not accounting for the upload rate and, therefore, the total communication rate of the scheme. This was done since the previous literature on this subject, \cite{ravi2018mmult}, \cite{kakar}, and \cite{koreans}, all used the download rate as their measure of performance. We will now see that both the download rate and the upload rate for polynomial codes both depend on the number, $N$, of servers.

Let $A \in \mathbb{F}_q^{r \times s}$ and $B \in \mathbb{F}_q^{s \times t}$. As in \eqref{partition1}, partition them as follows:
\begin{equation*}
    A = \begin{bmatrix}
    A_1 \\ \vdots \\ A_K
    \end{bmatrix},\quad 
    B = \begin{bmatrix} 
    B_1 & \cdots & B_L
    \end{bmatrix},\quad \text{so that}\quad 
    AB = \begin{bmatrix}
    A_1B_1 & \cdots & A_1B_L \\
    \vdots & \ddots & \vdots \\
    A_KB_1 & \cdots & A_KB_L
    \end{bmatrix}.
\end{equation*}
Thus, each $A_i \in \mathbb{F}_q^{\frac{r}{K} \times s}$ and each $B_i \in \mathbb{F}_q^{ s \times \frac{t}{L}}$. The random matrices will also belong, respectively, to these spaces.

Using a polynomial code a user will send a linear combination of the $A$'s and $R$'s and another one of the $B$'s and $S$'s to each server, requiring an upload of $rs/K + st/L$ symbols per server, for a total upload cost of $N(rs/K + st/L)$ symbols. Each server will then multiply the two matrices they received and send the user a matrix of dimensions $r/K \times t/L$, for a total download cost of $Nrt/KL$.  Thus, under our framework, minimizing the download, upload, or total communication costs are all equivalent to minimizing the number of servers, $N$. 

\update{A more thorough analysis on the communication and computational costs in SDMM can be found in \cite{notes}.}

\update{Let us conclude by briefly discussing the difference in total communication cost between $\mathsf{GASP}$ and the scheme of \cite{kakar}.  As we saw in Section \ref{workers}, for fixed $N$ and $T$ satisfying $T > N/6$ the download rates of these schemes are the same.  The scheme of \cite{kakar} achieves this rate by setting $L = 1$, while $\mathsf{GASP}$ achieves this rate by calculating the optimal value of $KL$, and choosing any values of $K$ and $L$ that yield this product.  For fixed values of $N$, $T$, and $KL$, minimizing the communication cost is equivalent to minimizing the upload cost $N(rs/K + st/L)$.  This is accomplished by choosing $K$ and $L$ to be as close to each other as possible, which $\mathsf{GASP}$ allows for.  In contrast, the scheme of \cite{kakar} which sets $L = 1$ ends up maximizing the upload cost subject to the given conditions.  For example, when $N = 20$ and $T = 6$, the scheme of \cite{kakar} sets $K = 4$ and $L = 1$, while $\mathsf{GASP}$ sets $K = L = 2$.  This results in a $20\%$ decrease in upload cost when $r = s = t$.}

%

\end{document}